\documentclass[twocolumn,aps,prc,floatfix]{revtex4}
\usepackage{graphicx}
\usepackage{dcolumn}
\usepackage{bm}
\usepackage{mhchem}
\usepackage{color}
\begin{document}

\title{Formation of large chunk of nuclear matter in heavy-ion collisions}

\author{Gao-Chan Yong$^{1,2}$}
\affiliation{$^{1}$Institute of Modern Physics, Chinese Academy of Sciences, Lanzhou 730000, China \\
$^{2}$School of Nuclear Science and Technology, University of Chinese Academy of Sciences, Beijing, 100049, China}

\begin{abstract}

In terrestrial laboratory with existing accelerators, the maximum multinucleon system produced so far is limited to that with total nucleon number being less than the sum of two colliding nuclei. Such situation may hinder our studies on the properties of nuclear many-body system. Here a way of producing large multinucleon system via multi-nucleus collision device Collider Plus in terrestrial laboratory is proposed. It is shown that large chunk of nucleonic matter can be produced through nuclear ternary fusion reactions at lower beam energies and also large block of dense matter can be formed via ternary collisions of heavy nuclei at intermediate energies. These investigations may shed lights on the studies of the synthesis of superheavy elements and the properties of bulk superdense nuclear matter.

\end{abstract}

\maketitle

\section{Introduction}
With six new synthetic elements with atomic numbers Z= 113-118 joining the periodic table, the 7th row of the periodic table of the elements is completed \cite{elem118}. The remaining specific questions frequently asked in the field of superheavy elements are, e.g., what are the heaviest nuclei that can exist? Where does the periodic table of elements end? \cite{sa19}. Traditionally, heavy-ion fusion reactions at energies around the Coulomb barrier have been used as a standard method to synthesize superheavy elements \cite{fu1,fu2}. Based on the mechanism of nuclear fusion-evaporation reactions, ion beams of typically and approximately single or double magic lighter nuclei, the projectiles, are accelerated and crash onto thin targets consisting of approximately single or double magic heavier nuclei. If the beam energy of the incoming ions is large enough to overcome the repulsive Coulomb forces between the nuclei of projectile ions and target nuclei, superheavy nuclei can be formed in complete fusion. Given the tiny probability of synthsizing superheavy nuclei, one needs a high intensity heavy-ion beam and a heavy actinide target of high stability as well as sophisticated target technology \cite{cc21}. On the other hand, there are strong theoretical suggestions that superheavy nuclei differ from lighter species because of their large charges counterbalancing the quantum shell effects. The presence of large electrostatic forces gives rise to strong Coulomb frustration effects in the nuclear system which challenges the current many-body theory \cite{np18}.

Ternary nuclear reactions involving three alpha nuclei were previous studies by Umar \emph{et al.} with the time-dependent Hartree-Fock (TDHF) theory. It was found in this study that during this reaction, a triple-alpha linear chain configuration of $^{12}C$ was formed, which subsequently made a transition to a triangular configuration before reaching near the ground state of $^{12}C$ \cite{triple10}. After this, the time-dependent density functional theory (TDDFT) was also used by Iwata \emph{et al.} to study heavy element production from ternary reactions of three $^{56}Fe$ \cite{ternary13}. Both studies are of interest in nucleosynthesis. To synthesize superheavy nuclei beyond Z = 118 by using hot fusion reactions with the double magic nucleus $^{48}$Ca projectile, one would need heavier (Z $>$ 98) targets. However, these heavier nuclei are usually both short lived and not available with sufficient amounts to perform fusion experiments. To avoid using heavier and short-lived targets, in the present study, an alternative method of synthesizing superheavy nuclei via a dynamical ternary fusion (e.g, $^{48}$Ca+$^{238}$U+$^{48}$Ca) is proposed. It is found that compound nuclei with large charge numbers are easily formed via multi-nucleus fusion reaction in terrestrial laboratory.

Another fundamental issue in nuclear physics is to explore the equation of state (EoS) of nuclear matter \cite{sci1,sci2,rmp17}, usually defined as its energy (or pressure) as a function of density. The EoS of dense matter is needed to calculate a series of properties of neutron stars. In particular, the knowledge of the EoS is necessary for the determination of the maximum mass M$_{max}$ of neutron stars (A compact object with its mass larger than M$_{max}$ would collapse into a black hole) \cite{ana}.
Because of the complicated nonperturbative feature of Quantum Chromodynamics (QCD), it is still a big challenge to determine the nuclear matter EoS from \emph{ab initio} QCD calculations, especially at suprasaturation densities \cite{epjc}. Therefore, to determine the EoS of dense matter from terrestrial heavy-ion collisions experiments is particularly important \cite{stock86}. In the heavy-ion collisions community, the investigation of constraints on the high-density behavior of nuclear matter has recently received new impetus from, e.g., CBM and HADES at GSI/FAIR, BM@N and MPD at NICA, CEE at HIAF, Dipole Hadron Spectrometer and Dimuon Spectrometer at J-PARC-HI, and STAR at RHIC BES-II \cite{qm2018}. To probe the phase transition boundary of dense QCD matter from the hadronic phase to the quark-gluon-plasma (QGP) phase in heavy-ion collisions, one needs to search for observabels that are sensitive to the stiffness of the EoS of dense matter \cite{cas2021}.

While the EoS of symmetric dense nuclear matter with an equal fraction of neutrons and protons has been reasonably constrained by analyzing the data on kaon production \cite{kaon1,kaon2} and
collective flow \cite{sci1} in heavy-ion collisions, one still cannot determine the critical density for the phase transition of dense matter from the hadronic phase to the QGP phase. Moreover, the EoS of dense \emph{neutron-rich} matter (i.e., neutron star matter) remains largely uncertain due to the poorly known high-density behavior of the isospin dependent part of nuclear matter EoS, characterized by the symmetry energy \cite{esym,esym2}. Both of them require a delicate determination of the EoS of dense matter. Therefore, one has to find more sensitive methods to probe the EoS of dense matter created in heavy-ion collisions. Besides searching for more EoS-sensitive observables such as the double strangeness $\Xi^{-}$ production \cite{cas2021}, it may be possible to create high sensitivity to the EoS via the production of long lived dense matter with large volume in heavy-ion collisions.

\section{methodology}
The present study is based on the framework of an isospin-dependent Boltzmann-Uehling-Uhlenbeck (BUU) transport model \cite{esym}, which is frequently used to describe the dynamical evolution of nuclear reaction. The BUU transport model describes the time evolution of the single-particle phase-space distribution function $f(\vec{r},\vec{p},t)$, which reads \cite{bertsch}
\begin{equation}
\frac{\partial f}{\partial
t}+\nabla_{\vec{p}}E\cdot\nabla_{\vec{r}}f-\nabla_{\vec{r}}E\cdot\nabla_{\vec{p}}f=I_{c},
\label{IBUU}
\end{equation}
where the phase-space distribution function $f(\vec{r},\vec{p},t)$ denotes the probability of finding a particle at time $t$ with momentum $\vec{p}$ at position $\vec{r}$. The left-hand side of Eq.~(\ref{IBUU}) denotes the time evolution of particle phase-space distribution function due to its transport and mean field, and the right-hand side collision item $I_{c}$ accounts for the modification of phase-space distribution function by elastic and inelastic two-body collisions. $E$ stands for a particle's total energy, which is equal to its instant kinetic energy $E_{kin}$ plus its corresponding single-particle potential $U$. The single-particle potential $U$ of a particle is given self-consistently by its
phase-space distribution function. Here the used Skyrme-type parametrization for the isoscalar term of the single-nucleon potential
gives the ground-state compressibility coefficient of nuclear matter $K$ = 230 MeV \cite{eos1,rmp17}. As for the isospin-dependent EoS, a density-dependent symmetry energy $E_{sym}= 32(\rho/\rho_0)^{\gamma}$ with $\gamma=1,3$ is used \cite{gamma99}. The choice of the symmetry energy parameter $\gamma$ is for the convenience of specifying its stiffness at high densities. Collisions of nucleons in projectile and target are modeled as that in Ref~\cite{esym}. For nuclear ternary reactions, in the simulations, one nucleus as target is fixed while the other two nuclei as projectiles collide with the static target nucleus in opposite directions. Dynamic evolution of the nuclear ternary reaction is same as that of the nuclear binary reaction.

\begin{figure}[thb]
\centering
\vspace{-0.7cm}
\includegraphics[width=0.48\textwidth]{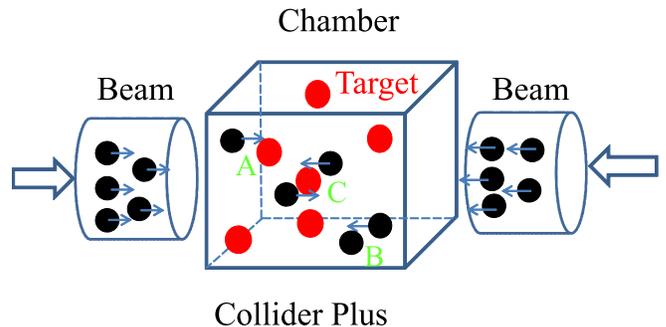}
\vspace{-0.5cm}
\caption{Sketch of the Collider Plus of nuclear ternary collision. Case A (fixed-target mode) and case B (conventional Collider mode) denote normal nuclear binary collisions while case C (Collider Plus mode) stands for the nuclear ternary collision. Two beams in opposite directions meet in the chamber in which target nuclei are pre-placed.} \label{collider}
\vspace{0.25cm}
\end{figure}
Fig.~\ref{collider} shows the sketch of designed device for the nuclear ternary collision Collider Plus. The nuclear binary collision Collider pattern has been extensively used nowadays, such as the eminent Relativistic Heavy Ion Collider (RHIC) at Brookhaven National Laboratory (BNL) in the US or the well known Large Hadron Collider (LHC) at the European particle physics laboratory CERN, i.e., European Council for Nuclear Research, in Switzerland. As a conjectured devise, two beams travel in opposite directions and collide with a static target in a chamber detector. There is no static target for conventional nuclear binary collision Collider. The third participant nucleus is pre-placed in the chamber as fixed target and the other two projectiles collide with it in opposite directions. Of course, the probability of the nuclear ternary collision is far below that of the nuclear binary collisions. If the reaction probability for the fixed-target mode is $P_{1}$ and the reaction probability of conventional Collider mode is $P_{2}$, the reaction probability $P$ for the new devised nuclear ternary collision mode (Collider Plus) reads
\begin{equation}\label{xsec}
  P=P_{1}\cdot P_{2}.
\end{equation}
Although $P$ is far below $P_{1}$ or $P_{2}$, the probability of the nuclear ternary collision is in principle experimentally feasible.

\section{Results and Discussions}
\begin{figure}[t]
\centering
\vspace{-0.5cm}
\includegraphics[width=0.48\textwidth]{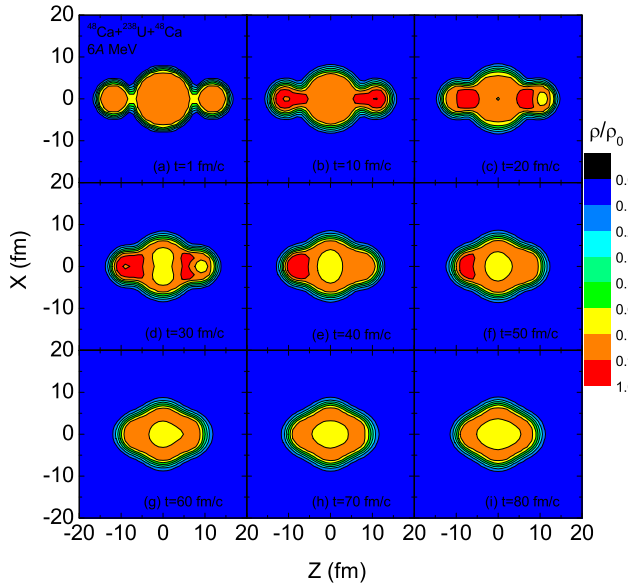}
\caption{Evolution of density contour in the X-Z plane of the nuclear ternary central $^{48}$Ca+$^{238}$U+$^{48}$Ca reaction with a beam energy of 6$\emph{A}$ MeV.} \label{contourxz}
\vspace{0.25cm}
\end{figure}
To study the synthesis of superheavy nuclei in terrestrial laboratory, the nuclear ternary central $^{48}$Ca+$^{238}$U+$^{48}$Ca reaction with a beam energy of 6$\emph{A}$ MeV, which is somewhat higher than the Coulomb barrier \cite{bass74}, is simulated within the framework of BUU transport model. In practice, one usually uses hot or cold fusion reactions \cite{back14,kag18}, such as Ca+$^{238}$U with the double magic nucleus $^{48}$Ca, to synthesize superheavy nuclei in the laboratory. To explore alternative avenue leading to the synthesis of superheavy nuclei, multi-nucleus fusion reaction to form large chunk of nucleonic matter is carried out. From the density contour in the X-Z plane in Fig.~\ref{contourxz}, it is seen that a larger compound nucleus is formed after two projectiles overcome the fusion or Coulomb barrier. It is also seen that the stability of the formed compound nucleus is achieved after t = 60 fm/c. In practice, the configuration of formed compound nucleus at t = 200 fm/c is almost the same as that at t = 60 fm/c which is not shown here. As discussed in Ref~\cite{np18}, the configuration of formed superheavy nuclei is expected to exhibit exotic topologies of nucleonic densities, such as voids (bubbles) or tori. Due to larger Coulomb effects for large $Z$ matter, semi-bubble structures may be common in many cases \cite{sa19}.

\begin{figure}[tbh]
\centering
\vspace{0.cm}
\includegraphics[width=0.48\textwidth]{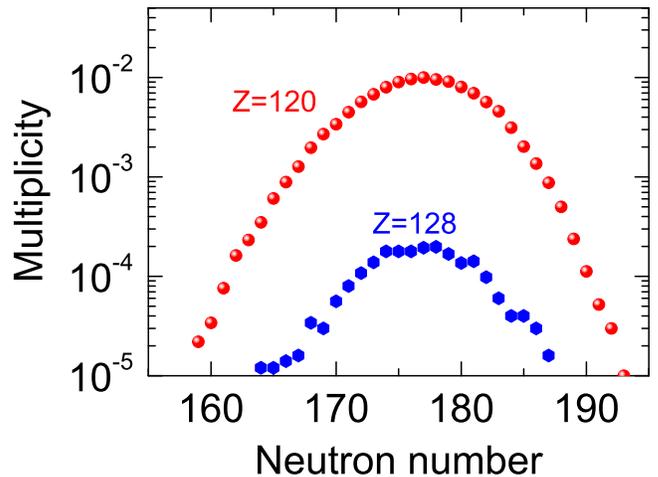}
\caption{Multiplicity of primevally produced isotopes Z = 120 and Z = 128 at t = 100 fm/c as a function of neutron number in the central $^{48}$Ca+$^{238}$U+$^{48}$Ca reaction with the beam energy of 6$\emph{A}$ MeV.} \label{frags}
\vspace{0.25cm}
\end{figure}

To shown more clearly the production of superheavy nuclei in the nuclear ternary fusion reaction $^{48}$Ca+$^{238}$U+$^{48}$Ca with the beam energy of 6 MeV per nucleon, the yields of superheavy isotopes Z = 120 and Z = 128 are demonstrated in Fig.~\ref{frags}. Because most BUU-type transport models are incapable of forming dynamically realistic nuclear fragments, certain types of afterburners, such as statistical and coalescence models, are normally used as a remedy. For the purpose of present exploration, a simple phase-space coalescence model \cite{coa1} is used, where a physical fragment is formed as a cluster of nucleons with relative momenta smaller than P$_{0}$ and relative distances smaller than R$_{0}$. The results presented here are obtained with P$_{0}$ = 260 MeV/c and R$_{0}$ = 3 fm (decided by deuteron radius, Fermi momentum at saturation and uncertainty relationship of quantum mechanics). From Fig.~\ref{frags}, it is seen that there are about 10$^{-2}$ isotope (Z = 120, N = 178) and about 10$^{-4}$ isotope (Z = 128, N = 178) productions. As noted in the simulations, the synthesized compound nucleus can exist beyond 30 fm/c $\simeq$ 10$^{-22}$ s and thus can be considered a nucleus \cite{lifetime}.

The property of dense nuclear matter is crucial to the study of many phenomena in heaven, such as that relevant to neutron stars and supernova \cite{shen98,james2000}. The property or stiffness of dense nuclear matter is also crucial for studying the phase transition of nuclear matter from the hadronic matter to the QGP matter \cite{Hung95}. To study the properties of dense matter, besides using neutron stars as natural laboratory in heaven \cite{gw17,fat18,esym2}, one frequently uses heavy-ion collisions to form compressed dense nuclear matter in terrestrial laboratory and then probes its properties via emitted final-state particles \cite{stock86,esym}.

The transient existence of dense matter formed in heavy-ion collisions causes difficulties in probing the properties of dense matter. Most observables, especially those suffering from strong final-state interactions with surrounding matter, in fact probe weighted-average properties of surrounding matter ranging from the dense to the dilute state. Also the small volume of dense matter shortens the time of interactions between the dense matter and a specific observable, thus decreases the sensitivity of the observable to the property of dense matter. To enhance the sensitivity, one of the most crucial factors is to increase the volume of produced dense matter in heavy-ion collisions. In this sense, multi-nucleus collisions and the formation of large chunk of dense matter are extremely useful for studying the properties of dense nuclear matter in terrestrial laboratory.

\begin{figure}[thb]
\centering
\vspace{-0.5cm}
\includegraphics[width=0.48\textwidth]{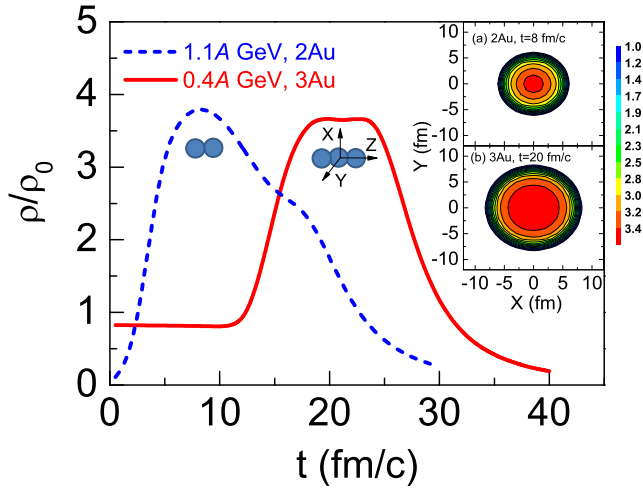}
\caption{Time evolution of central density reached in head-on nuclear binary Au+Au collision with a beam energy of 1.1 GeV per nucleon and head-on nuclear ternary Au+Au+Au collision with a beam energy of 0.4 GeV per nucleon, respectively. The inset shows density contour in the X-Y plane in the 2Au and 3Au collisions at maximum compression moments t=8 fm/c and 20 fm/c. All the plots are
depicted in the center of mass of system.} \label{dens}
\vspace{0.25cm}
\end{figure}
Fig.~\ref{dens} shows the production of dense matter in head-on 2Au and 3Au collisions as a function of time with beam energies of 1.1 and 0.4 GeV per nucleon, respectively. To match the maximum densities reached in 2Au and 3Au collisions, 0.4 and 1.1 GeV per nucleon beam energies are selected. While with almost the same maximum compression density, the lifetime and volume of the formed dense matter varies from 2Au to 3Au system. For the head-on nuclear binary Au+Au collision, the lifetime of produced dense matter with densities above 3.4 times saturation density ($\rho\geq3.4\rho_{0}$) lasts about 5 fm/c. While this changes to about 8 fm/c for the head-on nuclear ternary Au+Au+Au collision, which is about 60\% longer than that of nuclear binary Au+Au collision. Also, as shown in the inset of Fig.~\ref{dens}, the volume of dense matter formed in the nuclear ternary Au+Au+Au collision is several times larger than that in the nuclear binary Au+Au collision. Larger volume and longer lifetime of formed dense matter in the nuclear ternary collision would definitely have huge influence on the emission of final-state particles.

The density dependence of the nuclear symmetry energy is crucial for understanding not only the structure of rare isotopes and heavy-ion reactions, but also many interesting issues in astrophysics \cite{epja2014,jpg2014}. To determine the symmetry energy and thus the EoS of neutron-rich nuclear matter has been a long-standing goal of both nuclear physics and astrophysics \cite{rmp1974,ann1979}. To this end, enormous progress has been made in the constraints on the density-dependent symmetry energy, especially its high-density behavior \cite{ppnp2016}. The main challenge of probing the symmetry energy in heavy-ion collisions is the small sensitivities of all the symmetry-energy-sensitive observables (usually 10\% or so, due to the small strength of the symmetry potential relative to the isospin-independent part of the single-nucleon potential). The small sensitivity of observables thus frequently causes large deviations in interpreting experimental data by using diverse transport models \cite{guo14}. The larger volume and longer lifetime of formed dense matter in heavy-ion collisions may significantly enhance the sensitivities of the symmetry-energy-sensitive observables.

\begin{figure}[thb]
\centering
\vspace{0.cm}
\includegraphics[width=0.48\textwidth]{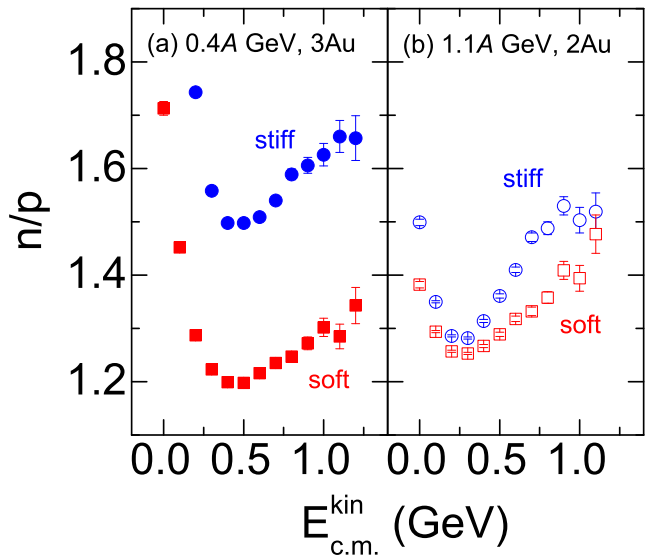}
\caption{Effects of the symmetry energy on the free neutron to proton ratio n/p at mid-rapidity $|y_{c.m.}|\leq0.1$ as a function kinetic energy in the central Au+Au+Au collisions with the beam energy of 0.4 GeV per nucleon and in the central Au+Au collisions with the beam energy of 1.1 GeV per nucleon, respectively. The density-dependent symmetry energy has a form of $E_{sym}= 32(\rho/\rho_0)^{\gamma}$ with parameter $\gamma$ being 1 and 3 for soft and stiff energies, respectively. All the analyses are conducted in the center of mass of system.} \label{rnp}
\vspace{0.25cm}
\end{figure}
Fig.~\ref{rnp} shows the sensitivity of the free neutron to proton ratio n/p to the high-density symmetry energy as a function of kinetic energy in 2Au and 3Au collisions with beam energies of 1.1 and 0.4 GeV per nucleon, respectively. It is clearly demonstrated that the sensitivity of the free neutron to proton ratio n/p to the symmetry energy in 3Au system is about 3 times that in 2Au system. The larger volume and longer lifetime of formed dense matter in 3Au system (shown in Fig.~\ref{dens}) enhances the effect of the EoS of dense matter on the symmetry-energy-sensitive observable free neutron to proton ratio n/p. The large sensitivity of observables has great significance for constraining the density dependence of the symmetry energy.

\section{Conclusions}
To summarize, based on a Collider Plus device that conducts nuclear ternary collisions, nuclear ternary reactions are studied. It is found that the nuclear ternary reaction with the beam energy above the Coulomb barrier can lead to the formation of extremely superheavy compound nuclei which may further form unknown superheavy nuclei via deexcitation. At higher beam energies, nuclear ternary collisions can produce dense matter with large volume and long lifetime, which can help us better understand the properties of dense nuclear matter. The study of the production of large chunk of nuclear matter in heavy-ion collisions at terrestrial laboratory could shed lights on the explorations of the properties of many-nucleon system. Since both the synthesis of superheavy nuclei and the study of the EoS of dense matter as well as other topics relevant to many-nucleon systems are of great significance in modern physics, it is essential to carry out such exploration program.

\section{Acknowledgments}
This work is Supported by the Strategic Priority Research Program of Chinese Academy of Sciences, Grant No. XDB34030000 and the National Natural Science Foundation of China under Grant No. 11775275.

\end{document}